# Low temperature magnetic structure of geometrically frustrated SrHo$_2$O$_4$

O Young[1], L C Chapon[2,3] and O A Petrenko[1]

[1] Department of Physics, University of Warwick, Coventry CV4 7AL, United Kingdom
[2] ISIS Facility, Rutherford Appleton Laboratory, Chilton, Didcot OX11 0QX, United Kingdom
[3] Institut Laue-Langevin, 6 Jules Horowitz, BP156, 38042 Grenoble Cedex 9, France

E-mail: o.young@warwick.ac.uk

**Abstract.** SrHo$_2$O$_4$ is a geometrically frustrated compound because the Ho$^{3+}$ ions are arranged in a network of triangles and hexagons and coupled with antiferromagnetic exchange interactions. Powder neutron diffraction measurements show that long-range magnetic order is established at $T_N = 0.68$ K. The low temperature magnetic structure is made up of two distinct components: a **k** = 0 long-range structure as well as a shorter-range low-dimensional structure with scattering intensity around the (0 0 $\frac{1}{2}$) position. The diffuse component of the magnetic scattering from SrHo$_2$O$_4$ remains very intense at temperatures well above $T_N$.

## 1. Introduction

Geometric frustration occurs in magnetic systems that cannot satisfy all the competing interactions because of their underlying lattice structure [1]. The Kagome [2], pyrochlore [3] and garnet [4] structures are well known frustrated systems and their magnetic ions are arranged in lattices that are made up of either corner- or edge-sharing triangles or tetrahedra. Interest in these systems comes from the rich variety of magnetic behaviour that is seen due to the influence of small perturbations (such as single ion anisotropy) and fluctuations (thermal or quantum) that will stabilise a ground state out of the macroscopic degeneracy of the classical Hamiltonian. The honeycomb lattice - made up of a planar network of hexagons - has become the subject of much theoretical interest since it has the lowest possible coordination number in two dimensions and thus quantum fluctuations are expected to play a big role in establishing its ground state [5]. Bipartite lattices, such as the honeycomb, are not expected to be frustrated, however they can be if next-nearest-neighbour exchange terms are not negligible.

Recently, a number of compounds based on the honeycomb lattice have been realised experimentally in $S = \frac{3}{2}$ systems Bi$_3$Mn$_4$O$_{12}$(NO$_3$) [6] and $\beta$-CaCr$_2$O$_4$ [7], as well as the classical spin systems Ba$Ln_2$O$_4$ [8] and Sr$Ln_2$O$_4$ [9]. Previous bulk property measurements of Sr$Ln_2$O$_4$ compounds (which crystallize in the CaFe$_2$O$_4$-type structure [10], space group $Pnam$) have revealed a discrepancy between the high Curie-Weiss constants [11] and the low Néel ordering temperatures [12, 13]. For one of the members of this family, SrEr$_2$O$_4$, we have already established the magnetic structure that appears below $T_N = 0.75$ K [12], despite having $\theta_{CW} = -13.5$ K [11]. We have now begun an extended investigation into the effects of frustration as a function of the rare earth element in the Sr$Ln_2$O$_4$ systems, firstly by studying SrHo$_2$O$_4$.

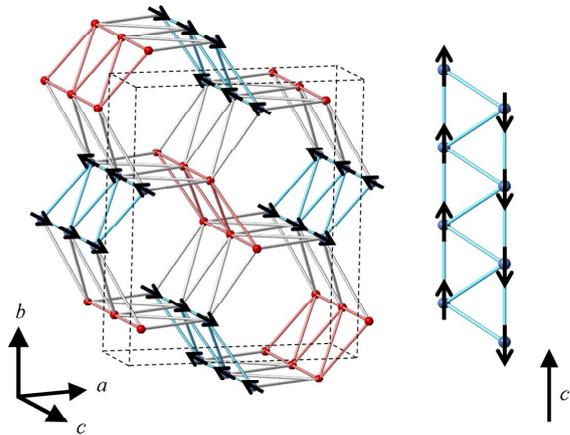

**Figure 1.** (Colour online) Magnetic structure of SrHo$_2$O$_4$ at $T = 45$ mK determined from a Rietveld refinement performed using the neutron scattering data collected on GEM (ISIS). When viewed in the $a$-$b$ plane, honeycombs of the Ho$^{3+}$ ions are visible. Triangular chains running along the $c$-axis connect the honeycomb layers and give rise to geometric frustration. The two crystallographically inequivalent Ho$^{3+}$ sites are shown in different colours as only one (blue) carries a significant magnetic moment of about 5 $\mu_B$ (the second Ho$^{3+}$ has no more than 0.2 $\mu_B$). All of the moments point along the $c$-axis, with neighbouring chains coupled antiferromagnetically. Swapping the moments between the red and blue sites has a negligible impact on the quality of the refinement.

From powder neutron diffraction data, we have established that the long-range ordered magnetic structure of SrHo$_2$O$_4$ (shown in figure 1) appears at $T_N = 0.68$ K. The temperature dependence of the magnetic scattering shows that in addition to the **k** = 0 Néel phase there exists some diffuse scattering, around the $(0\ 0\ \frac{1}{2})$ and symmetry related positions, that persists at temperatures well above $T_N$. In a related, isostructural compound, SrEr$_2$O$_4$, the coexistance of two magnetic phases well below $T_N$ [14] is thought to arise because one of the two crystallographically inequivalent Er$^{3+}$ sites orders in a long-range three-dimensional structure, whereas the other Er$^{3+}$ partially orders in a short-range magnetic phase.

## 2. Experimental Details

The SrHo$_2$O$_4$ sample was prepared from high purity starting compounds SrCO$_3$ and Ho$_2$O$_3$, in an off-stoichiometric ratio 1 : 0.8775. The powders were ground together and heated in air at 1400°C for a total of 48 hours (in an alumina crucible) with one intermediate grinding to ensure homogeneity of the mixture. The purity of the SrHo$_2$O$_4$ powder was checked by performing a Rietveld refinement using x-ray diffraction data, and this indicated the final purity to be greater than 99 %. Magnetic susceptibility measurements, on powder SrHo$_2$O$_4$, show a cusp at 0.7 K and this suggests the presence of a phase transition. For a single crystal of SrHo$_2$O$_4$ we have previously reported $T_N = 0.62$ K [13] from low temperature susceptibility measurements. We believe that the difference in $T_N$, between the powder and the single crystal SrHo$_2$O$_4$, may be due to slight non-stochiometry of the samples resulting from the crystal growth process. The GEM instrument at ISIS (Rutherford Appleton Laboratory, UK) was used to collect neutron scattering data in a range of temperatures from 45 mK to 10 K. The FULLPROF software suite was employed to refine the magnetic structure of SrHo$_2$O$_4$ using the powder data.

## 3. Results and Discussion

The proposed magnetic structure of SrHo$_2$O$_4$ that appears at $T_N = 0.68$ K was refined from powder neutron scattering data, and is shown in figure 1. The moments are aligned along the $c$-axis, since no $(00l)$ reflections were observed. The moments are coupled ferromagnetically in chains running along the $c$ direction, with neighbouring chains in the same ladder coupled antiferromagnetically. Only one of the crystallographically inequivalent sites of the Ho$^{3+}$ ions carries a significant magnetic moment of 5 $\mu_B$, the second has no more than 0.2 $\mu_B$. Swapping the moments between the sites in our model has a negligible impact on the quality of the refinement.

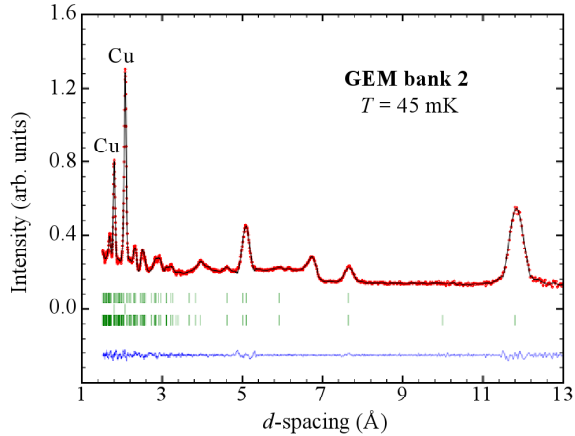

**Figure 2.** (Colour online) SrHo$_2$O$_4$ powder neutron diffraction refinement using the data collected in detector bank 2 on the GEM instrument at $T = 45$ mK. Data points and fit are shown in the upper curve, the difference between the observed and calculated pattern is shown in the lowest curve. Three sets of vertical tick marks (from top to bottom) correspond to the expected positions for nuclear peaks from SrHo$_2$O$_4$, reflections from the copper sample container and magnetic peaks from SrHo$_2$O$_4$ respectively. The peaks coming from the copper sample holder are labelled for clarity.

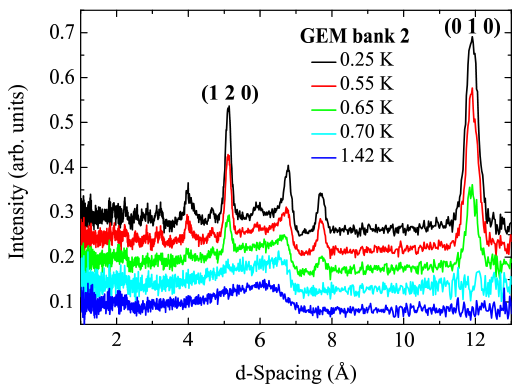

**Figure 3.** (Colour online) Temperature evolution of the magnetic scattering of SrHo$_2$O$_4$ (curves have been offset for clarity). Below 0.7 K magnetic Bragg peaks can be seen indicating the presence of long-range order. Above $T_N$, however, short-range correlations around the $(0\ 0\ \frac{1}{2})$ position are also observed. The magnetic peak due to these short-range correlations gets sharper (but does not change position) upon decreasing the temperature. The magnetic diffraction pattern is isolated by subtracting a 10 K background from the low temperature runs.

The FULLPROF refinement was performed using data collected in detector banks 2 to 5 of the GEM instrument simultaneously. Figure 2 only shows the magnetic structure refinement of the data collected in detector bank 2 at 45 mK, since in this detector bank all of the low $Q$ magnetic reflections from SrHo$_2$O$_4$ are visible. The diffuse scattering seen around the $(0\ 0\ \frac{1}{2})$ position in reciprocal space was treated as part of the background in the structure refinement as the limited information from powder data does not allow for a complete solution of this component of the magnetic scattering. The agreement factor for the magnetic phase in the detector bank shown in figure 2 is $R_{\mathrm{mag}} = 1.32$ %.

Figure 3 shows five neutron diffraction patterns collected in the temperature range 0.25 K to 1.42 K. Broad, diffuse scattering peaks are seen at temperatures much higher than $T_N$ (for example see the bottom pattern in figure 3 which was taken at 1.42 K). Bragg peaks begin to appear below 0.7 K, indicating the onset of long-range order. The peaks get more intense upon lowering the temperature, but in reciprocal space their positions remain the same. This long-range magnetic structure was found to be commensurate with the lattice and it can be indexed with the propagation vector $\mathbf{k} = 0$. The diffuse scattering that coexists with the long-range order, however, has the characteristic powder profile of a low-dimensional magnetic structure.

Within the refined magnetic structure there is a large difference between the values of the Ho$^{3+}$ moments that sit on crystallographically inequivalent sites. We believe that this is due to the coexistance of two separate magnetic structures in this material. The first is the long-range ordered three-dimensional magnetic structure (discussed above) that appears at $T_N = 0.68$ K, see figure 4a); the second is a short-range ordered low-dimensional structure that develops well above $T_N$, see figure 4b). It is useful to note that the magnetic scattering shown in figure 4b), at $T = 1.42$ K, is practically identical to the magnetic scattering shown in figure 16b) in the paper by Karunadasa *et al.* [11], where the data were obtained by subtracting the 5.05 K background from the data collected at 1.72 K. Further neutron scattering experiments on single crystals of SrHo$_2$O$_4$, which are already available [15], are now needed to give us greater insight into the nature of the unusual diffuse scattering features seen in the powder data.

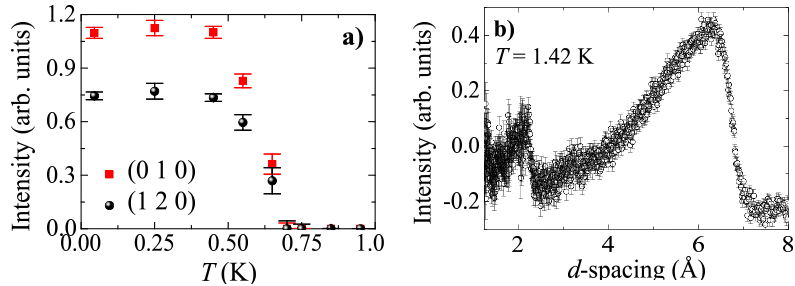

**Figure 4.** a) Temperature dependence of the integrated magnetic intensity for the two strongest magnetic peaks: (0 1 0) and (1 2 0), and b) magnetic scattering above $T_N$, at $T = 1.42$ K isolated by subtracting a 10 K background from data collected at 1.42 K.

## 4. Conclusions

We have determined the magnetic structure of SrHo$_2$O$_4$ which forms below $T_N = 0.68$ K. The powder neutron scattering data have revealed the coexistance of two magnetic phases in the same material: a fully ordered commensurate phase with the propagation vector $\mathbf{k} = 0$, and a short-range ordered, low-dimensional phase which persists at temperatures much higher than $T_N$. Further neutron scattering experiments on single crystals of SrHo$_2$O$_4$ are planned to fully determine the nature of the diffuse scattering seen at low temperatures.